\documentclass[a4paper,fleqn,useAMS,usenatbib,usedcolumn]{mnras}

\usepackage{psfrag,color,epsfig,rotating}
\usepackage{float}

\usepackage{graphicx}	
\usepackage{amsmath}	
\usepackage{amssymb}	
\usepackage{multicol}	
\usepackage{bm}		
\usepackage{pdflscape}	
\usepackage{tabularx}  
\usepackage{caption}

\usepackage[T1]{fontenc}
\usepackage{ae,aecompl}

\def\HI{{\ion{H}{i}}~}
\def\HII{{\ion{H}{ii}}~}
\def\xb{\bar{x}_{\rm \ion{H}{i}}}
\def\Tb{{T_{\rm b}}}
\def\tTb{\tilde{T}_{\rm b}}
\def\bTb{\bar{T}_{\rm b}}
\def\k{\mathbfit{k}}
\def\x{\mathbfit{x}}
\def\cov{\mathbfss{C}}
\def\dcov{\mathbfss{c}}
\def\pi{\upi}

\def\p1{Paper~I}

\title[The evolution of the power spectrum error-covariance]{Statistics 
of the epoch of reionization (EoR) 21-cm signal -- II. The evolution of the 
power spectrum error-covariance}

\author[R. Mondal, S. Bharadwaj and S. Majumdar]{Rajesh Mondal,$^{1,2}$\thanks{E-mail: \href{mailto:rm@phy.iitkgp.ernet.in}{rm@phy.iitkgp.ernet.in}} 
Somnath Bharadwaj$^{1, 2}$ and  Suman Majumdar$^3$\\
$^1$ Department of Physics, Indian Institute of Technology Kharagpur, Kharagpur -- 721302, India\\
$^2$ Centre for Theoretical Studies, Indian Institute of Technology Kharagpur, Kharagpur -- 721302, India\\
$^3$ Department of Physics, Blackett Laboratory, Imperial College, London SW7 2AZ, UK}

\date{}
\pubyear{2016}

\begin{document}
\label{firstpage}
\pagerange{\pageref{firstpage}--\pageref{lastpage}}
\maketitle


\begin{abstract}
The EoR 21-cm signal is expected to become highly non-Gaussian as 
reionization progresses. This severely affects the error-covariance 
of the EoR 21-cm power spectrum which is important for predicting the 
prospects of a detection with ongoing and future experiments. Most 
earlier works have assumed that the EoR 21-cm signal is a Gaussian 
random field where (1) the error-variance depends only on the power 
spectrum and the number of Fourier modes in the particular $k$ bin, 
and (2) the errors in the different $k$ bins are uncorrelated. Here 
we use an ensemble of simulated 21-cm maps to analyze  the 
error-covariance at various stages of reionization. We find that 
even at the very early stages of reionization ($\xb \sim 0.9 $) the 
error-variance significantly exceeds the Gaussian predictions at small 
length-scales ($k > 0.5 \,{\rm  Mpc}^{-1}$) while they are consistent 
at larger scales. The errors in most $k$ bins (both large and small 
scales), are however found to be correlated. Considering the later 
stages ($\xb=0.15$), the error-variance shows an excess in all $k$ 
bins within $k \ge 0.1 \, {\rm Mpc}^{-1}$, and it is around $200$ times 
larger than the Gaussian prediction at $k \sim 1 \, {\rm Mpc}^{-1}$. 
The errors in the different $k$ bins are all also highly correlated, 
barring the two smallest $k$ bins which are anti-correlated with the 
other bins. Our results imply that the predictions for different 21-cm 
experiments based on the Gaussian assumption underestimate the errors, 
and it is necessary to incorporate the non-Gaussianity for more 
realistic predictions.  
\end{abstract}


\begin{keywords}
methods: statistical -- cosmology: theory -- dark ages, reionization, 
first stars -- large-scale structure of Universe -- diffuse radiation. 
\end{keywords}


\section{Introduction}
\label{sec:intro}
The Epoch of Reionization (EoR) is an important period in the history
of our universe. During this era the hydrogen in our universe
gradually changed its state from being neutral (\ion{H}{i}) to ionized
(\ion{H}{ii}). Our knowledge in this regard is guided so far by two
major observational constraints. The measurements of the Thomson
scattering optical depth of the cosmic microwave background (CMB)
photons from the free electrons (e.g.
\citealt{komatsu11,planck15,planck16,planck_tau16} etc.) and the observations of
the Lyman-$\alpha$ absorption spectra of the high redshift quasars
(e.g. \citealt{becker01,fan03,white03,goto11,becker15} etc.). These
observations together suggest that this epoch probably extended over a
wide redshift range $6 \lesssim z \lesssim 12$ (e.g.
\citealt{mitra11,mitra13,mitra15,robertson13,robertson15,planck16} etc.).
However, most of the fundamental issues associated with this epoch,
such as the properties of sources contributing to the ionization
photon budget, gradual evolution in their properties, topology of the
brightness temperature maps at different stages of the reionization
etc. remain unknown till date.

The brightness temperature fluctuations of the 21-cm signal
originating due to the hyperfine transition in the neutral hydrogen
has the potential to probe this rather complex epoch.  There is
considerable effort underway to detect the EoR 21-cm signal through
radio interferometry using e.g. the
{GMRT\footnote{\href{http://www.gmrt.ncra.tifr.res.in}{http://www.gmrt.ncra.tifr.res.in}}}
\citep{paciga13},
{LOFAR\footnote{\href{http://www.lofar.org}{http://www.lofar.org}}}
\citep{yatawatta13,haarlem13},
{MWA\footnote{\href{http://www.haystack.mit.edu/ast/arrays/mwa}{http://www.haystack.mit.edu/ast/arrays/mwa}}
  \citep{tingay13,bowman13,dillon14}} and
{PAPER\footnote{\href{http://eor.berkeley.edu}{http://eor.berkeley.edu}}}
\citep{parsons14,jacobs14,ali15}. Apart from these first generation
radio interferometers, the detection of this signal is also one of the
key science goals of the future radio telescopes such as the
{SKA\footnote{\href{http://www.skatelescope.org}{http://www.skatelescope.org}}}
{\citep{mellema13,koopmans15}} and
{HERA\footnote{\href{http://reionization.org}{http://reionization.org}} 
\citep{furlanetto09}}.

In the recent past a significant amount of effort has gone into in
making quantitative predictions of the sensitivity of different
ongoing and upcoming experiments to measure the 21-cm signal through
its power spectrum (e.g. \citealt{morales05, mcquinn06, beardsley13}
etc.). The majority of these studies have been done by simulating the
expected EoR 21-cm signal and incorporating the specific telescope
response to it (e.g.  \citealt{thomas09, jensen13, pober14, majumdar16} etc.).

The redshifted 21-cm signal is largely dominated by different
foregrounds which are $\sim 4-5$ orders of magnitude stronger than the
signal \citep{ali08,bernardi09,ghosh12,pober13,moore15}.  Modelling
these foregrounds accurately and removing them from the actual data is
the major obstacle in detecting the cosmological 21-cm signal. In this
work, we assume that the foregrounds can be removed accurately, and we
only focus entirely on the signal.

Power spectrum estimation for the EoR 21-cm signal is restricted by
statistical uncertainties inherent to the measurements. A part of
these uncertainties arises due to the system noise  which
is Gaussian, and depends on the instrument and the total observation time. 
It may happen that in some observations the system noise dominates 
the total error budget. However the system noise can, in principle, be 
made arbitrarily small by increasing the  observation  time and  
we do not consider the system noise in the present work. 
Apart from the system noise, any statistical
cosmological signal always comes with an intrinsic uncertainty which
is known as the \emph{cosmic variance}. While taking into account the
effect of this cosmic variance in their measurements, all of the
above mentioned sensitivity estimation studies have effectively
assumed that the system noise and the EoR 21-cm signal are both
independent Gaussian random variables.  This is possibly a valid
assumption as long as reionization is at its early stages and one is
looking at sufficiently large scale power of the signal. During the
early stages of the EoR, the \HI is expected to trace the underlying
matter distribution. As reionization proceeds, creation and growth of
the ionized regions introduces non-Gaussianity \citep{bharadwaj05a,
  mondal15} in the \HI field. The simple assumption of the signal to
be a Gaussian random variable for power spectrum error estimates does
not hold.  It is necessary to quantify the uncertainties correctly to
obtain a proper interpretation of the signal.

\citet{mondal15} have first studied how the non-Gaussianities present
in the \HI distribution during the EoR affect the error predictions
for the 21-cm power spectrum. If the signal were a Gaussian random
field, one would expect the signal-to-noise ratio (SNR) to scale as
the square-root of the number of independent measurements. Using a
large ensemble of statistically independent realizations of EoR 21-cm
signal, they have shown that for a fixed observational volume, it is
not possible to achieve a SNR above a certain limiting value $[{\rm
  SNR}]_l$, even if one increases the number of independent Fourier
modes that goes into the estimation of the power spectrum of the
signal. They have also demonstrated that the non-Gaussianity present
in the EoR 21-cm signal, which is mainly driven by the distribution of
the \HII regions, gradually increases with the increasing volume and
number of these \HII regions as reionization progresses. At a fixed
redshift $z = 8$, they have considered different values of global
neutral fraction $\xb$ and shown that the value of $[{\rm SNR}]_l$
falls with the progress in reionization. The non-Gaussian effects play
an important role in the error predictions for the EoR 21-cm power
spectrum particularly in the later stages of the EoR.

Building up on the work of \citet{mondal15}, in \citet{mondal16}
(hereafter \p1) we presented a detailed and generic theoretical frame
work to interpret and quantify the effect of non-Gaussianity on the
error estimates for the 21-cm power spectrum through the
error-covariance $\cov_{ij}$. Using the analytic calculation,
we have identified two sources of contribution to the $\cov_{ij}$. One
is the usual variance for a Gaussian random field and the other is the
non-Gaussian component which is quantified through the trispectrum. We
validated this theoretical frame work using a large ensemble of
simulated EoR 21-cm signal at a fixed neutral fraction of $\sim 0.5$,
for two different simulation volumes. This analysis established the
fact that, due to the non-Gaussianity, not only the different Fourier
modes $\tTb (k)$ but also the errors in the power spectrum estimated
in different $k$ bins, are correlated. There are significant
correlations between the errors at the small length scales and between
the small and the large length scale. We also observed a small
anti-correlation between the errors in the smallest and the
intermediate $k$ values. It is important to note that the theoretical
frame work established and used in \citet{mondal15} and \p1 are very
generic and not limited only to the EoR 21-cm signal but can be
applied to the analysis of any non-Gaussian cosmological signal such
as the galaxy redshift surveys
\citep{hamilton06,neyrick08,neyrinck11,Harnois13,mohammed14,caron14}.

The entire analysis of \p1 was done at a fixed redshift
($=8$) and neutral fraction ($\approx 0.5$). It does not
address the issue of how the error-covariance evolves with the
evolving non-Gaussianity in the signal as reionization progresses. In
this paper, as a follow up of \p1, we study the evolution
of the error-covariance of the 21-cm power spectrum more 
generally with the evolving redshift and neutral fraction 
as ionization of the IGM progresses during the EoR. Here 
we identify the different sources that contribute to the 
non-Gaussianity and subsequently to the $\cov_{ij}$. We 
test the theoretical formalism for interpreting the 
error-covariance of the power spectrum presented in \p1 
for these evolving simulated \HI fields. We further 
identify the length scale ranges where the non-Gaussianity 
becomes significant (or in other words the trispectrum 
contribution to the error-covariance is significant) and 
how those evolve. We also try to investigate the causes of 
correlation and anti-correlation between the errors in the 
power spectrum estimated in different $k$ bins. 

The structure of this paper is as follows. In \autoref{sec:cov},
we briefly summarize the theoretical formalism presented in
\p1 for interpreting the error-covariance of the power
spectrum. We describe our semi-numerical simulations which are used to
generate the ensembles of the EoR 21-cm signal in \autoref{sec:sim}.  
In \autoref{sec:results}, we describe our results
i.e. the estimated power spectrum error-covariance at different stages
of reionization. Finally, in \autoref{sec:summary}, we discuss and
summarize our findings.

Throughout this paper, we have used the Planck+WP best fit values of
cosmological parameters $\Omega_{\rm m0}=0.3183$,
$\Omega_{\rm \Lambda0}=0.6817$, $\Omega_{\rm b0}h^2=0.022032$, $h=0.6704$,
$\sigma_8=0.8347$, and $n_{\rm s}=0.9619$ \citep{planck14}.

\section{The power spectrum error-covariance}
\label{sec:cov}
In this section we briefly summarize the theoretical formalism for
interpreting the EoR 21-cm power spectrum error-covariance $\cov_{ij}$.
The reader is refered to Sections 2 and 3 of \p1 for a detailed
discussion.  We consider $\tTb (\k)$ which is the Fourier transform
of the EoR 21-cm brightness temperature fluctuation
$\delta \Tb (\x)=\Tb (\x)- \bTb $ in a cubic comoving
volume $V$ with periodic boundary conditions.  The primary quantity of
interest here is the power spectrum $P(k)$ of the brightness
temperature fluctuations which  is defined as 
\begin{equation}
P(\k)=V^{-1} \langle \tTb(\k) \, \tTb(-\k) \rangle \,.
\label{eq:pkdef}
\end{equation}
We also introduce the trispectrum $T(\k_a,\k_b,\k_c,\k_d)$ which is defined
through 
\begin{align}
&\langle \tTb (a) \tTb (b) \tTb (c) \tTb (d) \rangle = V^2 [ \, \delta_{a+b,0}~\delta_{c+d,0}~P(a) P(c)
\nonumber \\
&+ \delta_{a+c,0}~\delta_{b+d,0}~P(a) P(b)
+ \delta_{a+d,0}~\delta_{b+c,0}~P(a) P(b)] 
\nonumber \\
&+ V \delta_{a + b + c + d,0}~T (a,b,c,d)
\label{eq:tk1}
\end{align}
where we have used the notation $\tTb (a) \equiv \tTb ({\k_a})$.
Note that the trispectrum is zero for a Gaussian random field.

Here we have considered the bin averaged power spectrum $\bar{P} ({k_i})$
evaluated in different bins in
$k$ space. The bins here are spherical shells of width $\Delta k_i$ in
Fourier space.  We use the binned power spectrum estimator, which for
the $i~$th bin is defined as
\begin{equation}
\hat{P}(k_i)= \frac{1}{N_{k_i} V} \sum_{\k}  \tTb (\k) \,\tTb (-\k) \, ,
\label{eq:est}
\end{equation}
where $\sum_{\k}$, $N_{k_i}$ and $k_i$ respectively refer to the 
sum, the number and the average comoving wavenumber of all the 
Fourier modes in the $i~$th bin. The ensemble average over many 
statistically independent realizations of $\tTb (\k)$ gives the 
mean bin-averaged power spectrum 
$\bar{P} ({k_i}) = \langle \hat{P}({k_i}) \rangle$.

The question of  interest here is -- `How accurately can the 
power spectrum be  estimated from a given EoR 21-cm data set?'.  
Any estimation of the EoR 21-cm power spectrum is constrained 
by the errors,  and here we only consider the statistical 
uncertainty which is inherent to the EoR 21-cm signal, the 
\textit{cosmic variance}. We quantify this uncertainty through 
the error-covariance of the binned power spectrum estimator
\begin{equation}
\cov_{ij} = \langle [ \hat{P}({k_i}) - \bar{P} ({k_i})]
[ \hat{P}({k_j}) - \bar{P} ({k_j})] \rangle\,.
\label{eq:cov0}
\end{equation}
%

The error-covariance $\cov_{ij}$ is found (\p1) to scale with 
the volume as $V^{-1}$,  and the  values of  $\cov_{ij}$ are also 
found to span a very large dynamical range across the $k$ values 
which we have considered here. It is more convenient to consider 
the  dimensionless form of the error-covariance matrix (equation 
33 of \p1) which is defined as
\begin{equation}
\dcov_{ij}=\frac{\cov_{ij} \, V \ k_i^{3/2} k_j^{3/2}}{(2 \pi)^2
\bar{P}(k_i) \,  \bar{P}(k_j)} \, . 
\label{eq:dcov}
\end{equation}
This does not have any explicit volume dependence, and it has been 
calculated (equation 34 of \p1) to be
\begin{equation}
\dcov_{ij}=A_i^2 \left(
\frac{k_i}{\Delta k_i} \right) \delta_{ij}+ t_{ij} \,.
\label{eq:cov}
\end{equation}
where the Kronecker delta $\delta_{ij}$ is $1$ if $i=j$ and $0$ 
otherwise. Here
\begin{equation}
A_i=\sqrt{\frac{\overline{P^2}(k_i)}{[\bar{P}(k_i)]^2}} \,.
\label{eq:Ai}
\end{equation}
where 
\begin{equation}
\overline{P^2}(k_i)=\frac{1}{N_{k_i}} \sum_{\k} P^2(\k) 
\end{equation}
is the square of the power spectrum  averaged over the $i$~th bin. 
The dimensionless quantity $A_i$ is a number of order unity introduced 
in \citet{mondal15}. The value  of $A_i$ is expected to vary from 
bin to bin. However, these variations are expected to be small, and 
we may expect a value $A_i \approx 1$ in most situations.

The dimensionless bin-averaged trispectrum (equation  35 of \p1)
is defined as 
\begin{equation}
t_{ij}=\frac{\bar{T} (k_i,k_j) \ k_i^{3/2} \, k_j^{3/2}  }{(2
  \pi)^2  \bar{P} (k_i) \,  \bar{P} (k_j)} \, ,
\label{eq:dtrisp}
\end{equation}
where 
\begin{equation}
\bar{T} (k_i,k_j) = \frac{1}{N_{k_i} N_{k_j}} 
\sum_{\k_a \in i, \k_b \in j} T (\k_a,-\k_a,\k_b,-\k_b)
\label{eq:tri}
\end{equation}
is the average trispectrum, $\k_a$ and $\k_b$ here are summed over the
$i~$th and the $j~$th bins respectively. The
trispectrum is zero for a Gaussian random field.
%


The diagonal elements of the error-covariance $\dcov_{ii}$
(equation~\ref{eq:cov}) quantify the variance of the error in the
estimated power spectrum. We have $\dcov_{ii} = A_i^2 ({k_i}/{\Delta
  k_i})$ if the EoR 21-cm signal is a Gaussian random field. In our
analysis we have used logarithmic binning which implies that $(k_i/
\Delta k_i) \approx 1.9$ is nearly constant across all the bins.  In
this case, we expect $\dcov_{ii}$ to have a nearly constant value
$\approx 2$ in all the $k$ bins.  The EoR 21-cm signal becomes
increasingly non-Gaussian as the IGM becomes more and more
ionized. This manifests itself as a non-zero contribution from the
trispectrum to $\dcov_{ii}$. As a consequence the value of
$\dcov_{ii}$ exceeds the value $\approx 2$ expected for a Gaussian
random field. We interpret this excess as arising from the trispectrum
which develops when the EoR 21-cm signal becomes non-Gaussian.

The off-diagonal terms of $\dcov_{ij}$ quantify the correlations
between the errors in the power spectrum estimated at different
bins. These terms are all zero {\it i.e.} the errors in the different
bins are uncorrelated if the EoR 21-cm signal is a Gaussian random
field. However, the EoR 21-cm signal becomes increasingly non-Gaussian
as reionization progresses, and we expect the off-diagonal terms to
develop non-zero values.  We interpret any statistically significant
non-zero off-diagonal component of $\dcov_{ij}$ as arising from the
trispectrum which develops when the EoR 21-cm signal becomes
non-Gaussian.

We have used equation~(\ref{eq:cov}) to interpret the results  in the
subsequent analysis of this paper.  We expect the diagonal elements 
$\dcov_{ii}$ to have a nearly constant value $\approx 2$ and the off-diagonal
elements of $\dcov_{ij}$ to be zero  if the EoR 21-cm signal is a Gaussian
random  field.  We have interpreted any deviations from these values as
arising from the trispectrum $t_{ij}$ which arises due to the non-Gaussian
nature of the EoR 21-cm signal. 

\section{Simulating the E\textsc{\lowercase{o}}R redshifted 21-\textsc{\lowercase{cm}} signal}
\label{sec:sim}
\subsection{The Signal Ensemble~(SE)}
\label{sec:simSE}
We have generated the redshifted EoR 21-cm signal using semi-numerical 
simulations which involve three main steps. First, we use a particle 
mesh $N$-body code to generate the dark matter distribution at the 
different redshifts listed in \autoref{tab:history}. We have run 
simulations with comoving volume $V = [215\, {\rm Mpc}]^3$ using 
$3072^3$ grids of spacing $0.07 \, {\rm Mpc}$ and a mass 
resolution of $1.09 \times10^8 M_{\sun}$. In the next step we use 
the Friends-of-Friends (FoF) algorithm \citep{davis85} to identify 
collapsed halos in the dark matter distribution. We have used a 
fixed linking length of $0.2$ times 
the mean inter-particle distance and also set the criterion that a 
halo should have at least $10$ dark matter particles which implies 
a minimum halo mass of $M_{\rm min}=1.09 \times10^9 M_{\sun}$.

\begin{table}
\centering
\caption{This tabulates the redshifts ($z$) and corresponding mass-averaged
neutral fraction ($\xb$) where we have simulation outputs for the 21-cm signal.}
\label{tab:history}
\begin{tabular}{cc}
\hline
\hline
\,$z$\, & \,$\xb$\,\rule{0pt}{2.6ex}\rule[-1.2ex]{0pt}{0pt}\\
\hline
13 & 0.98 \rule{0pt}{2.6ex}\rule[-1.2ex]{0pt}{0pt}\\
11 & 0.93 \rule{0pt}{2.6ex}\rule[-1.2ex]{0pt}{0pt}\\ 
10 & 0.86 \rule{0pt}{2.6ex}\rule[-1.2ex]{0pt}{0pt}\\
9 & 0.73 \rule{0pt}{2.6ex}\rule[-1.2ex]{0pt}{0pt}\\
8 & 0.50 \rule{0pt}{2.6ex}\rule[-1.2ex]{0pt}{0pt}\\
7 & 0.15 \rule{0pt}{2.6ex}\rule[-1.2ex]{0pt}{0pt}\\
\hline
\end{tabular}
\end{table}


The third and final step generates the ionization map based on the 
excursion set formalism of \citet{furlanetto04b}. It is assumed that 
the hydrogen traces the dark matter and the halos host the sources
which emit the ionizing radiation. It is also assumed that the number 
of ionizing photons emitted by a source is proportional to the mass 
of its host halo with the  constant of proportionality 
being quantified through a dimensionless parameter $N_{\rm ion}${\footnote
{One should note here that several unknown degenerate parameters of 
reionization such as the star formation efficiency of the early 
galaxies, their UV photon production efficiency and the escape 
fraction of the UV photons from these galaxies are all combined 
together in the parameter $N_{\rm  ion}$.}}.
In addition to $N_{\rm  ion}$, the simulations have another free parameter 
$R_{\rm mfp}$ the mean free path of the ionizing photons. We determine 
whether a grid point is ionized or not by smoothing the hydrogen density 
field and the photon density field using spheres of different radii 
starting from $R_{\rm min}$ (the grid size) to $R_{\rm mfp}$. A particular 
grid point is considered to be ionized if for any smoothing radius the 
photon density exceeds the hydrogen density at that grid point. The 
simulated \HI distribution is then mapped to the redshift space using 
the peculiar velocities of the simulation particles to generate 21-cm 
brightness temperature maps. Our semi-numerical simulations closely 
follow \citet{choudhury09b} and \citet{majumdar14} to generate the ionization field, and the 
resulting field is mapped on to redshift space following the methodology 
of \citet{majumdar13}. The final ionization maps and the 21-cm brightness 
temperature maps are generated on a grid that is eight times coarser 
than the one used for the $N$-body simulations.
%

\begin{figure}
\psfrag{xh1}[c][c][1][0]{\Large $\xb$}
\psfrag{z}[c][c][1][0]{\Large $z$}
\psfrag{10}[c][c][1][0]{\Large $10$}
\centering
\includegraphics[width=.47\textwidth]{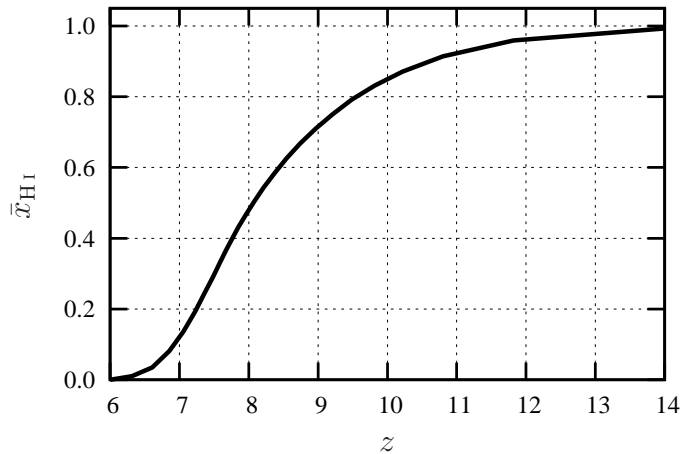}
\caption{This shows the reionization history ($\xb$ as a function of
  $z$) that we have obtained from our simulations.}
\label{fig:history}
\end{figure}

%
It is possible to achieve different reionization histories $\xb(z)$ by
varying the parameters $N_{\rm ion}$ and $R_{\rm mfp}$, $\xb(z)$ here
refers to the mass-averaged \HI fractions. The redshift evolution of
$\xb(z)$ during EoR is largely unknown, apart from the fact that the
\HI reionization ended before a redshift of $6$ (e.g.
\citealt{becker01,fan03,becker15}) and the total integrated Thomson
scattering optical depth has a value $\tau = 0.058 \pm 0.012$
\citep{planck16}. We have set $R_{\rm mfp} = 20\,{\rm Mpc}$ which is
consistent with the findings of \citet{songaila10} from the study of
Lyman limit systems at low redshifts.  The value of $N_{\rm ion}$ was
fixed (at $23.21$) so as to achieve $50\%$ ionization at $z=8$, and
the resulting reionization history is shown in
\autoref{fig:history}. We see that reionization is complete by $z \sim
6$ in the resulting reionization history, further we have $\tau=0.057$
which is consistent with the measured optical depth. The values of
$\xb$ obtained from our simulation are tabulated as a function of $z$
in \autoref{tab:history}.

\begin{figure*}
\centering
\includegraphics[width=1.015\textwidth, angle=0]{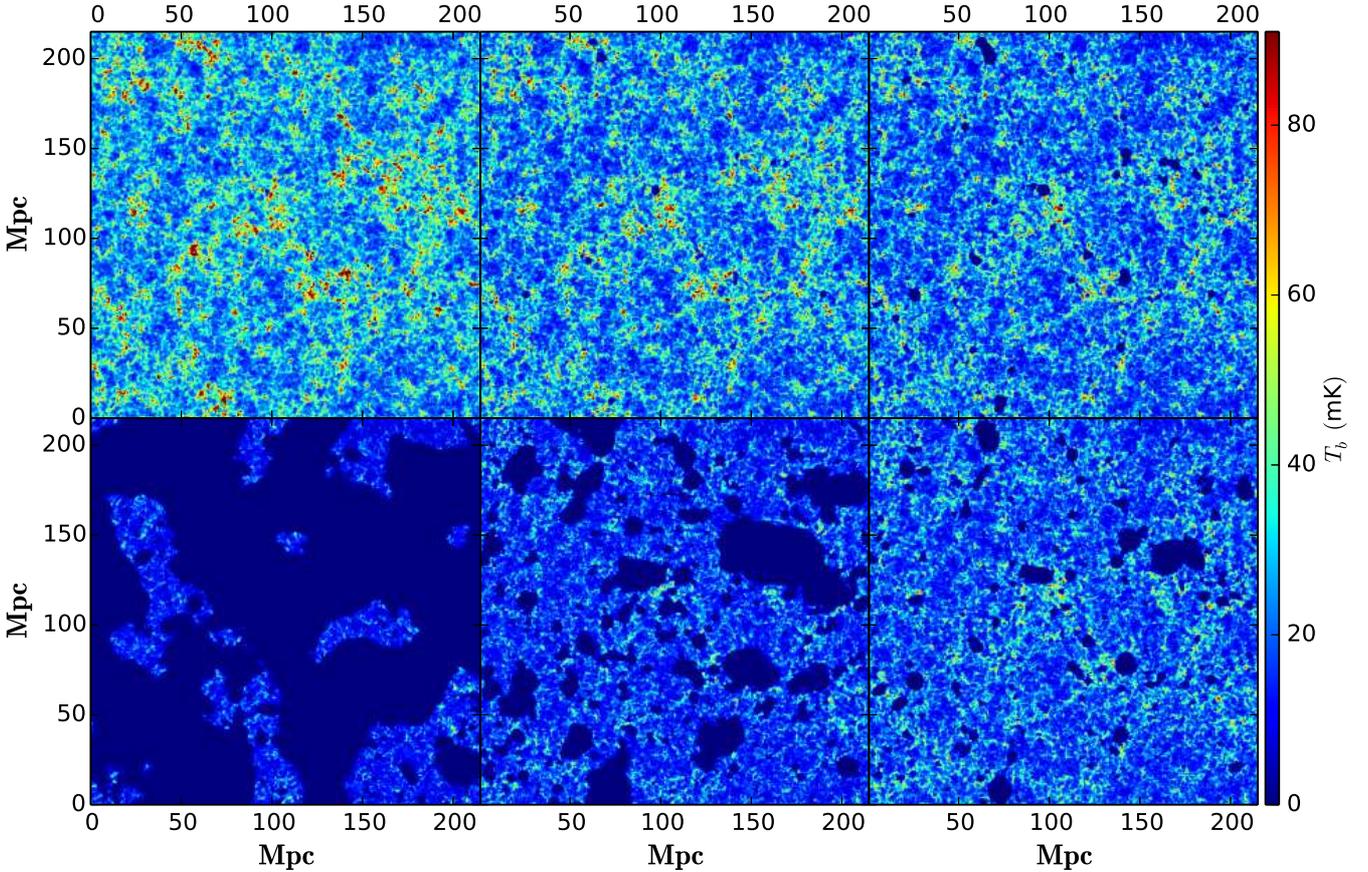}
\caption{This shows one realization of the two dimensional sections
  through the simulated \HI brightness temperature maps for the $[z,\,
  \xb]$ values tabulated in \autoref{tab:history}. In these panels,
  the value of $\xb$ decreases in the clockwise direction starting
  from the top-left panel. Volume of these simulation boxes are
  $[215\, {\rm Mpc}]^3$. These maps are in redshift space, where the
  line of sight direction is along the vertical axis.}
\label{fig:HI_map}
\end{figure*}

We have run $50$ independent  realizations of the simulations to 
generate an ensemble of $50$ statistically independent realizations of 
the EoR 21-cm signal
%
%
We refer to this ensemble as the Signal Ensemble~(SE). In summary, the
Signal Ensemble (SE) contains $50$ statistically independent
realizations of the 21-cm signal for each of the redshifts $z$ and
mass-averaged neutral fractions $\xb$ listed in
\autoref{tab:history}. \autoref{fig:HI_map} shows two dimensional
sections through one realization of the signal for all the $\xb$
values listed in \autoref{tab:history}. We have used the SE to
estimate the bin-averaged power spectrum $\bar{P} ({k_i})$ and the
error-covariance $\cov_{ij}$ at different stages of the
reionization.
%

\subsection{The Randomized Signal Ensemble~(RSE)}
\label{sec:simRSE}
We use the Randomized Signal Ensemble~(RSE) to interpret the diagonal
terms of the error-covariance $\cov_{ij}$ or equivalently the
dimensionless error-covariance $\dcov_{ij}$.  As noted earlier in
\autoref{sec:cov} we expect $\dcov_{ii} = A_i^2 ({k_i}/{\Delta k_i} )$
if the EoR 21-cm signal is a Gaussian random field. We interpret any
excess relative to this prediction as arising from the trispectrum
$t_{ii}$ which arises when the EoR 21-cm signal becomes
non-Gaussian. Here we have used logarithmic binning which implies that
$(k_i/ \Delta k_i)\approx 1.9$ is nearly constant across all the
bins. The difficulty is that it is not possible to predict the precise
value of $A_i$. It is straight forward to see (equation~\ref{eq:Ai})
that we have a constant value $A_i=1$ if the power spectrum $P(\k)$
has exactly the same value at all the Fourier modes $\k$ in the
bin. However, this simple assumption is obviously violated, for
example redshift space distortion causes the value of $P(\k)$ to vary
depending on the direction of $\k$.  Further, we have no prior idea of
how the simulated EoR power spectrum $P(\k)$ varies across the
different Fourier modes within a bin. We have overcome this problem by
constructing the RSE as briefly discussed below. The reader is
referred to the section 5.1 of \p1 for a detailed discussion on the
RSE.

Each realization of the  Randomized Signal Ensemble~(RSE) is a mixture 
of Fourier modes  $\tTb (\k)$ drawn from all $50$ realizations of the 
Signal Ensemble (SE). Considering a particular realization of the RSE, 
the modes which originate from different realizations of the signal are 
uncorrelated. This ensures  that the trispectrum, which quantifies the 
correlation between the signal at different Fourier modes, is at least 
$50$ times smaller for the RSE as compared to the SE. The actual 
suppression of the trispectrum $t_{ii}$ depends on the number of modes in 
the $i~$th bin, and our earlier studies show that the actual suppression 
can be considerably more than a factor of $50$ (Figure~5 of \p1). In 
this work, we  have assumed that $t_{ii} \approx 0$ for the RSE which 
allows us to interpret  $[\dcov_{ii}]_{\rm RSE}$ entirely in terms of 
the power spectrum. Further, the signal is randomized in such a way 
that we expect both $\bar{P}(k_i)$ and  $\overline{P^2}(k_i)$, and 
therefore also $A_i$ to have exactly the same values for the RSE as 
compared to the SE. The RSE therefore gives a direct estimate of the 
error-covariance  that would be expected if the trispectrum were zero 
i.e. 
\begin{equation}
[\dcov_{ii}]_{\rm RSE} = A_i^2  ({k_i}/{\Delta k_i} ) \,.
\label{eq:dcovrese}
\end{equation}
In the subsequent analysis we have used the difference 
\begin{equation}
[\dcov_{ii}]_{\rm SE} - [\dcov_{ii}]_{\rm RSE} =t_{ii}
\label{eq:tii}
\end{equation}
to directly estimate the reduced trispectrum. 

\subsection{Ensemble of Gaussian Random Ensembles~(EGRE)}
\label{sec:EGRE}
We expect the off-diagonal terms of $\dcov_{ij}$ to be zero for 
a Gaussian random field. However, we cannot straightaway interpret 
the non-zero off-diagonal terms in $\dcov_{ij}$ estimated from SE 
as arising from non-Gaussianity in the EoR 21-cm signal because 
the SE has a finite number of realizations. To appreciate this 
we consider a Gaussian Random Ensemble~(GRE) which has the same  
number of realizations of the 21-cm signal as the SE. The 
off-diagonal terms of the error-covariance $[\dcov_{ij}]_{\rm G}$ 
estimated from GRE will not be zero but will have random 
fluctuations around zero due to the limited number of 
realizations. To determine the statistical significance of 
$\dcov_{ij}$, one needs to compare it against the random 
fluctuation of $[\dcov_{ij}]_{\rm G}$. We have used $50$ 
independent GREs to construct an Ensemble of Gaussian 
Random Ensembles~(EGRE) following the idea presented in 
Section 5.2 of \p1. The EGRE has been used to estimate 
the variance $[\delta \dcov_{ij}]_{\rm G}^2$ of $[\dcov_{ij}]_{\rm G}$. 
We have compared the estimated $\dcov_{ij}$ against 
$[\delta \dcov_{ij}]_{\rm G}$ to determine whether our results 
are statistically significant or not.

\section{Results}
\label{sec:results}
\autoref{fig:HI_map} shows two dimensional sections through one
realization of our simulated three-dimensional 21-cm maps at the
redshifts listed in \autoref{tab:history}. We see that the \HI very
closely follows the underlying dark matter field during the early
stages of reionization (top three panels of \autoref{fig:HI_map}). We expect 
that the brightness temperature is, to a good approximation, a Gaussian random
field at these early stages ($\xb = 0.98,\,0.93$ and $0.86$). However, 
the non-linearities in the matter density at small scales
introduces some  amount of non-Gaussianity even  during this early phase
of reionization. The `inside-out' reionization implemented in these
simulations implies that the high-density regions get ionized first
and the low-density regions later. Small \HII bubbles located at the
high density peaks can be seen even at the early stages. These \HII
bubbles grow in both size and in number, and gradually overlap at the
later stages of reionization. Finally, at the last stage of
reionization we have small islands of \HI in an almost completely
ionized IGM. In summary, we see that as reionization proceeds the 21-cm
signal undergoes a transition from a state where it primarily traces
the nearly Gaussian matter fluctuations to a phase where it is nearly
entirely determined by a few discrete \HI regions which survive to the
end of the EoR. We do not show a visualization of the RSE or GRE here,
The readers are referred to Figure 1 of \p1 for a visual impression of
the RSE and the GRE.
%

\begin{figure}
\psfrag{pk}[c][c][1][0]{\large ${\Delta^2_{\rm b}}\, \, ({\rm mK})^2$}
\psfrag{k}[c][c][1][0]{\large $k\, \, ({\rm Mpc}^{-1}$)}
\psfrag{10}[c][c][1][0]{\large $10$}
\psfrag{0.1}[c][c][1][0]{\large $0.1$}
\psfrag{1.0}[c][c][1][0]{\large $1$}
\psfrag{z=13, xh1=0.98}[c][c][1][0]{\large $z=13$,\,$\xb=0.98$\,\,\,\,\,\,\,\,}
\psfrag{11,         0.93}[c][c][1][0]{\large $11$,\hspace{1.1cm}$0.93$\,\,\,\,}
\psfrag{10,         0.86}[c][c][1][0]{\large $10$,\hspace{1.1cm}$0.86$\,\,\,\,}
\psfrag{9,         0.73}[c][c][1][0]{\large $9$,\hspace{1.1cm}\,$0.73$\,\,\,\,}
\psfrag{8,         0.50}[c][c][1][0]{\large $8$,\hspace{1.1cm}\,$0.50$\,\,\,\,}
\psfrag{7,         0.15}[c][c][1][0]{\large $7$,\hspace{1.1cm}\,$0.15$\,\,\,\,}
\centering
\includegraphics[width=0.481\textwidth]{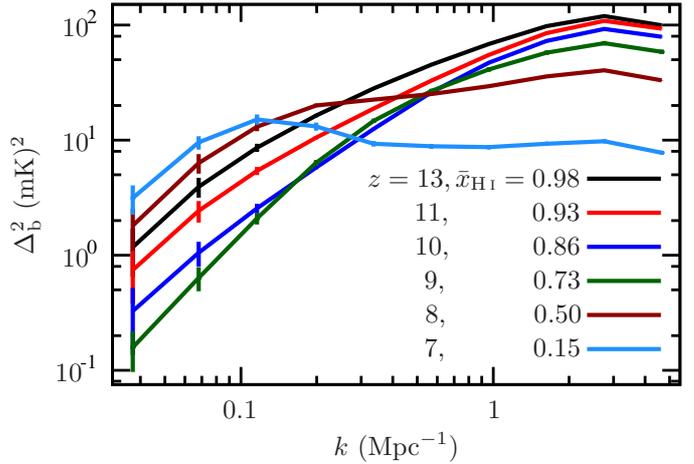}
\caption{This shows the mean squared 21-cm brightness temperature
  fluctuations $\Delta_{\rm b}^2(k)$ and its $1\sigma$ error bars
  (estimated from $50$ different realizations of the simulated signal)
  for the $[z,\, \xb]$ values shown in the figure.}
\label{fig:pk}
\end{figure}

%
\autoref{fig:pk} shows the mean square brightness temperature
fluctuations $\Delta_{\rm b}^2(k) = k^3\,P(k)/2 \pi^2$ of the
redshifted 21-cm signal as a function of $k$ for the different values
of $z$ and $\xb$ considered here.  For each redshift we have estimated
the average power spectrum $\bar{P}(k)$ and its $1\sigma$ errors
$\delta P(k) = \sqrt{\cov_{ii}}$ using the $50$ statistically
independent realizations of SE. We have divided the $k$ range into
$10$ equally spaced logarithmic bins for these estimates. We first
discuss the power spectrum from the early stages of reionization,
i.e. the redshifts $z = 13,\,11$ and $10$. During these stages, at
large length-scales $k \le 0.2 \, {\rm Mpc}^{-1}$ the shape of the
21-cm power spectrum remains nearly the same as the underlying matter
power spectrum.  However, the amplitude of $\Delta_{\rm b}^2(k)$
slowly decreases as the redshift decreases in contrast to the matter
power spectrum whose amplitude increases with decreasing
redshift. This is due to the fact that reionization preferentially
wipes out the highest density regions in the `inside-out' scenario
implemented here. The decrease in the amplitude of the power spectrum
is less at small scales where we also notice a steepening of the power
spectrum. The decrease in the large-scale amplitude slows down at $z
=9$, and we have a reversal in this trend at the lower redshifts.  As
reionization proceeds from $z=8$ to $7$ the amplitude of the power
spectrum increases at large scales $k < 0.1\,{\rm Mpc}^{-1}$, however
the amplitude decreases at small scales where the power spectrum
becomes nearly flat.  At these redshifts the turnaround in
$\Delta_{\rm b}^2(k)$ roughly occurs at a $k$ which corresponds to the
radius of the typical ionized regions. Our results here are consistent
with several previous studies performed using a variety of different
simulation techniques (e.g. \citealt{mcquinn07,lidz08,barkana09,choudhury09b, 
  mesinger11, jensen13, majumdar13, iliev14, majumdar16, dixon16}
etc.).
%

\begin{figure*}
\psfrag{cov}[c][c][1][0]{\Large $\dcov_{ii}$}
\psfrag{k}[c][c][1][0]{\large $k\, \, ({\rm Mpc}^{-1})$}
\psfrag{SE}[c][c][1][0]{\large SE}
\psfrag{RSE}[c][c][1][0]{\large RSE}

\psfrag{z=13, xh1=0.98}[c][c][1][0]{\large $z=13$, $\xb=0.98$}
\psfrag{z=11, xh1=0.93}[c][c][1][0]{\large $z=11$, $\xb=0.93$}
\psfrag{z=10, xh1=0.86}[c][c][1][0]{\large $z=10$, $\xb=0.86$}
\psfrag{z=9, xh1=0.73}[c][c][1][0]{\large $z=9$, $\xb=0.73$}
\psfrag{z=8, xh1=0.50}[c][c][1][0]{\large $z=8$, $\xb=0.50$}
\psfrag{z=7, xh1=0.15}[c][c][1][0]{\large $z=7$, $\xb=0.15$}

\psfrag{10}[c][c][1][0]{\large $10$} \psfrag{0.1}[c][c][1][0]{\large
  $0.1$} \psfrag{1.0}[c][c][1][0]{\large $1$} \psfrag{0.1
  1.0}[c][c][1][0]{\large $0.1$\hspace{1.7cm}\,\,\,$1$} \centering
\includegraphics[width=1.015\textwidth]{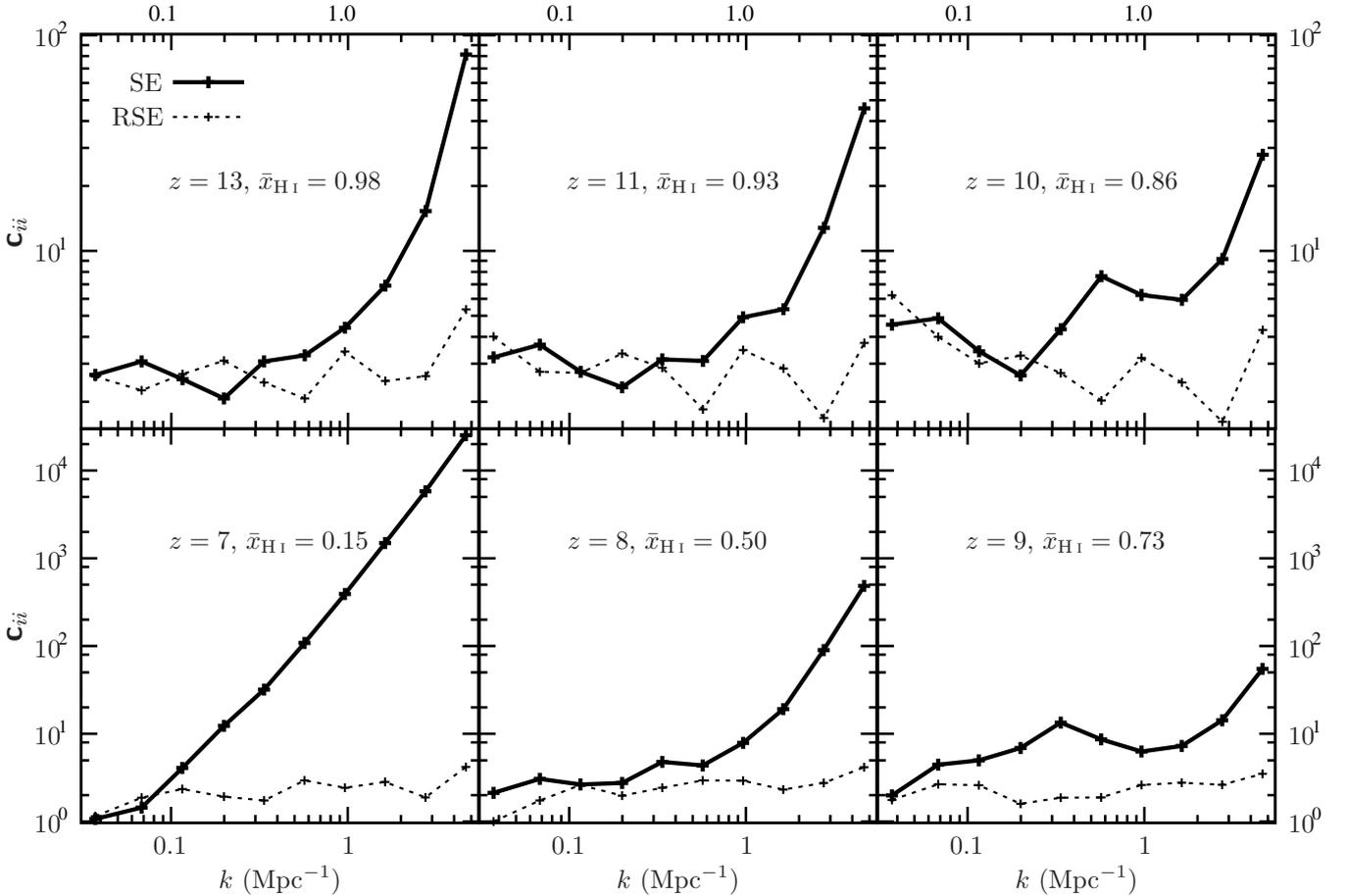}
\caption{This shows the dimensionless error-variance $\dcov_{ii}$
  for SE and RSE for the $[z,\, \xb]$ values shown in the figure.}
\label{fig:cov_SE_RSE}
\end{figure*}

%
We next turn our attention to the  error-covariance matrix $\cov_{ij}$, and 
we first consider the diagonal elements which give an estimate of the 
error-variance $[\delta P(k_i)]^2 = {\cov_{ii}}$ of the  estimated power 
spectrum. \autoref{fig:cov_SE_RSE} shows the diagonal elements of the 
dimensionless error-covariance matrix $\dcov_{ii}$  defined in 
equation~(\ref{eq:dcov}). Here we use $[\dcov_{ii}]_{\rm SE}$ to refer to the 
values which have been estimated from the $50$ realizations of the simulated 
EoR signal, namely the Signal Ensemble (SE) and we use equation~(\ref{eq:cov}) 
to interpret these results. In addition to this the figure also shows 
$[\dcov_{ii}]_{\rm RSE}$ which (equation~\ref{eq:dcovrese}) gives an estimate 
of the results that would be expected if the EoR  signal were a Gaussian 
random field with zero trispectrum $t_{ii}=0$. We see (\autoref{fig:cov_SE_RSE}) 
that the values of $[\dcov_{ii}]_{\rm RSE}$  are predominantly confined in the 
range $2$ to $4$ with a maximum value $[\dcov_{ii}]_{\rm RSE} \sim 6$ at the 
largest $k$ bin. In contrast to this, we find that the values of
$[\dcov_{ii}]_{\rm SE}$ vary over a much wider range with a maximum value
$[\dcov_{ii}]_{\rm SE} > 10^4$ at the  largest $k$ bin for $\xb=0.15$. 
We interpret any difference between $[\dcov_{ii}]_{\rm SE}$ and 
$[\dcov_{ii}]_{\rm RSE}$ as arising due to the non-Gaussianity of the EoR 
21-cm signal. 
We first discuss the results  for  the highest
redshift $z=13$  that we have considered here, shown in the  top left panel of
\autoref{fig:cov_SE_RSE}.
It is reasonably valid  to assume that at this stage of reionization
($\xb=0.98$) the \HI
distribution  directly traces the underlying  matter. The only
non-Gaussianities here are those that arise due to non-linear 
gravitational clustering  of the underlying matter distribution. We see
(\autoref{fig:cov_SE_RSE}) that the \HI signal is consistent with a Gaussian 
random field at large scales ($k \le 0.5 \, {\rm Mpc}^{-1}$) where 
$[\dcov_{ii}]_{\rm SE}$ is comparable to $[\dcov_{ii}]_{\rm RSE}$. The values of 
$[\dcov_{ii}]_{\rm SE}$, however, increase sharply with increasing $k$ for 
$k > 0.5 \, {\rm Mpc}^{-1}$ reaching a value $[\dcov_{ii}]_{\rm SE} \sim 10^2$ 
at the largest $k$ bin. The \HI signal here is non-Gaussian at
intermediate and small  length-scales ($k > 0.5 \, {\rm Mpc}^{-1}$) where
$[\dcov_{ii}]_{\rm SE}$ is  larger than $[\dcov_{ii}]_{\rm RSE}$.  The
non-Gaussianity seen here is nearly entirely due to the non-linear
gravitational clustering of the underlying matter distribution.
The results show a  similar behaviour at $z=11$ and $10$.
The effect of non-Gaussianity extends to relatively larger length-scales when 
the neutral fraction falls to $\xb=0.73$ at $z=9$ (\autoref{fig:cov_SE_RSE}). 
Here we see that $[\dcov_{ii}]_{\rm SE}$ is well in excess of
$[\dcov_{ii}]_{\rm RSE}$  for $k > 0.1 \, {\rm Mpc}^{-1}$, and this behaviour
possibly extends to 
even smaller values of $k$ all the way to $k \ge 0.04 \, {\rm Mpc}^{-1}$. The
behaviour become a little complicated for $\xb=0.5$ at $z=8$. Here, at small scales
the values of $[\dcov_{ii}]_{\rm SE}$ increase relative to those at $z=9$, the 
reverse however is found at intermediate scales $0.1\,< k < 1\,{\rm Mpc}^{-1}$. 
The results, however, are quite unambiguous for $\xb=0.15$ at $z=7$ where 
we find that $[\dcov_{ii}]_{\rm SE}$ is well in excess of $[\dcov_{ii}]_{\rm RSE}$ 
for $k \ge 0.1 \, {\rm Mpc}^{-1}$. Here the value of $[\dcov_{ii}]_{\rm SE}$ 
is found to increase monotonically with increasing $k$, and it has values
$\sim 10^2$ and $\sim 10^5$ at $k \sim 0.5 \, {\rm Mpc}^{-1}$ and 
$k \sim 5.0 \, {\rm Mpc}^{-1}$ respectively. We see that the effect 
of non-Gaussianity becomes particularly important across a wide range of
length-scales in the late stages of reionization.
%

\begin{figure*}
\psfrag{cov}[c][c][1][0]{\Large $\dcov_{ii}$}
\psfrag{xh1}[c][c][1][0]{\Large $\bar{x}_{\rm HI}$}
\psfrag{pk}[c][c][1][0]{\large ${\Delta^2_{\rm b}}\, \, ({\rm mK})^2$}

\psfrag{SE}[c][c][1][0]{\large SE}
\psfrag{RSE}[c][c][1][0]{\large RSE}

\psfrag{k=0.12Mpc-1}[c][c][1][0]{\large$k=0.12 \,{\rm Mpc}^{-1}$}
\psfrag{k=0.57Mpc-1}[c][c][1][0]{\large$k=0.57 \,{\rm Mpc}^{-1}$}
\psfrag{k=2.75Mpc-1}[c][c][1][0]{\large$k=2.75 \,{\rm Mpc}^{-1}$}

\psfrag{10}[c][c][1][0]{\large $10$}
\psfrag{4}[c][c][1][0]{\large $4$}
\psfrag{5}[c][c][1][0]{\large $5$}
\psfrag{6}[c][c][1][0]{\large $6$}
\psfrag{0.2}[c][c][1][0]{\large $0.2$}
\psfrag{0.4}[c][c][1][0]{\large $0.4$}
\psfrag{0.6}[c][c][1][0]{\large $0.6$}
\psfrag{0.8}[c][c][1][0]{\large $0.8$}
\psfrag{1.0}[c][c][1][0]{\large $1$}
\centering
\includegraphics[width=1.015\textwidth]{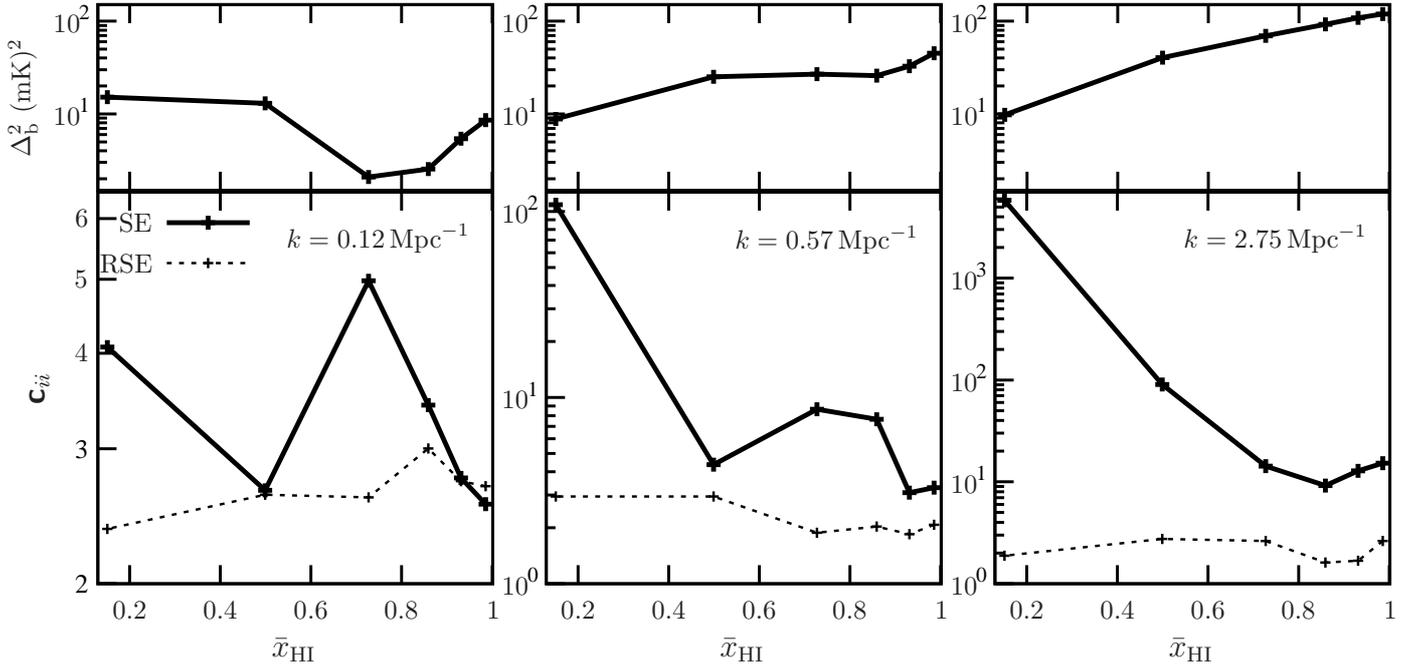}
\caption{This shows the dimensionless error-variance $\dcov_{ii}$
  for SE as a function of global neutral fraction $\bar{x}_{\rm HI}$
  for three representative wave numbers $k = 0.12,\, 0.57$ and
  $2.75\,{\rm Mpc^{-1}}$. }
\label{fig:cov_SE}
\end{figure*}

%
\autoref{fig:cov_SE} shows how $[\dcov_{ii}]_{\rm SE}$ and $[\dcov_{ii}]_{\rm RSE}$
at three different fixed values of $k$ evolve as a function of $\xb$.
This figure reinforces the findings highlighted in the previous
discussion. The values of $\Delta_{\rm b}^2(k)$ are also shown in the
corresponding top panels for reference. We see that at small
$(k=2.75 \, {\rm Mpc}^{-1})$ and intermediate  $(k=0.57 \, {\rm Mpc}^{-1})$
length-scales, the differences between $[\dcov_{ii}]_{\rm SE}$ and
$[\dcov_{ii}]_{\rm RSE}$ are clearly visible over the entire range of
evolution ($\xb=0.98$  to $0.15$) considered here.
This implies that at these length-scales
the 21-cm signal is significantly non-Gaussian 
even at the earliest stages of reionization $(\xb \sim~1)$.  As mentioned earlier,
initially the \HI directly traces the underlying matter distribution and
the non-Gaussianities here are entirely due to the non-linear gravitational
clustering of the underlying matter distribution. As reionization proceeds,
it preferentially wipes out the \HI in the non-linear high density peaks.
At small scales, this causes a drop in the non-Gaussianity of the \HI signal
in the early stages of reionization, and we see this reflected as a decrement
in $[\dcov_{ii}]_{\rm SE}$ as $\xb$ falls from $0.98$ to $0.86$. Subsequently,
the value of $[\dcov_{ii}]_{\rm SE}$ at small scales increases monotonically
by roughly three orders of magnitude as reionization proceeds and the value of 
$\xb$ drops from $0.86$ to $0.15$.
The non-Gaussianity here is largely dominated by the ionized regions.
The behaviour is more complicated at intermediate scales where
$[\dcov_{ii}]_{\rm SE}$ initially decreases, as reionization proceeds
the value of $[\dcov_{ii}]_{\rm SE}$
then increases till $\xb =0.73$,  dips again at $\xb=0.5$ and finally
increases at $\xb=0.15$. 
At large scales ($k= 0.12 \, {\rm Mpc}^{-1}$) we find that the values of
$[\dcov_{ii}]_{\rm SE}$ are comparable to those of $[\dcov_{ii}]_{\rm RSE}$ for
$\xb \ge 0.8$ indicating that at large scales the 21-cm signal is consistent
with a Gaussian random field in the early stages of reionization. The values
of $[\dcov_{ii}]_{\rm SE}$ are larger than those of $[\dcov_{ii}]_{\rm  RSE}$ for
$\xb = 0.73$ and $0.15$ indicating  that the effect of non-Gaussianity
extends to large scales as reionization proceed. However, the value of
$[\dcov_{ii}]_{\rm  SE}$ dips at $\xb=0.5$ where it is comparable to that of
$[\dcov_{ii}]_{\rm  RSE}$. This behaviour is similar to that seen at intermediate
scales where also the non-Gaussianity does not grow monotonically as
reionization proceeds but exhibits a dip at $\xb = 0.5$. Note that the 
peak value  $[\dcov_{ii}]_{\rm  SE} \sim 5$ at large scales is
considerably smaller than  the peak values of $[\dcov_{ii}]_{\rm  SE}$
at intermediate or small scales.

\begin{figure*}
\psfrag{k}[c][c][1][0]{\large $k\, \, ({\rm Mpc}^{-1})$}
\psfrag{cov}[c][c][1][0]{\large $r_{ij}$}
\psfrag{                                  k                          }[c][c][1][0]{\large $k\, \, ({\rm Mpc}^{-1}$)}
\centering
\includegraphics[width=1.015\textwidth]{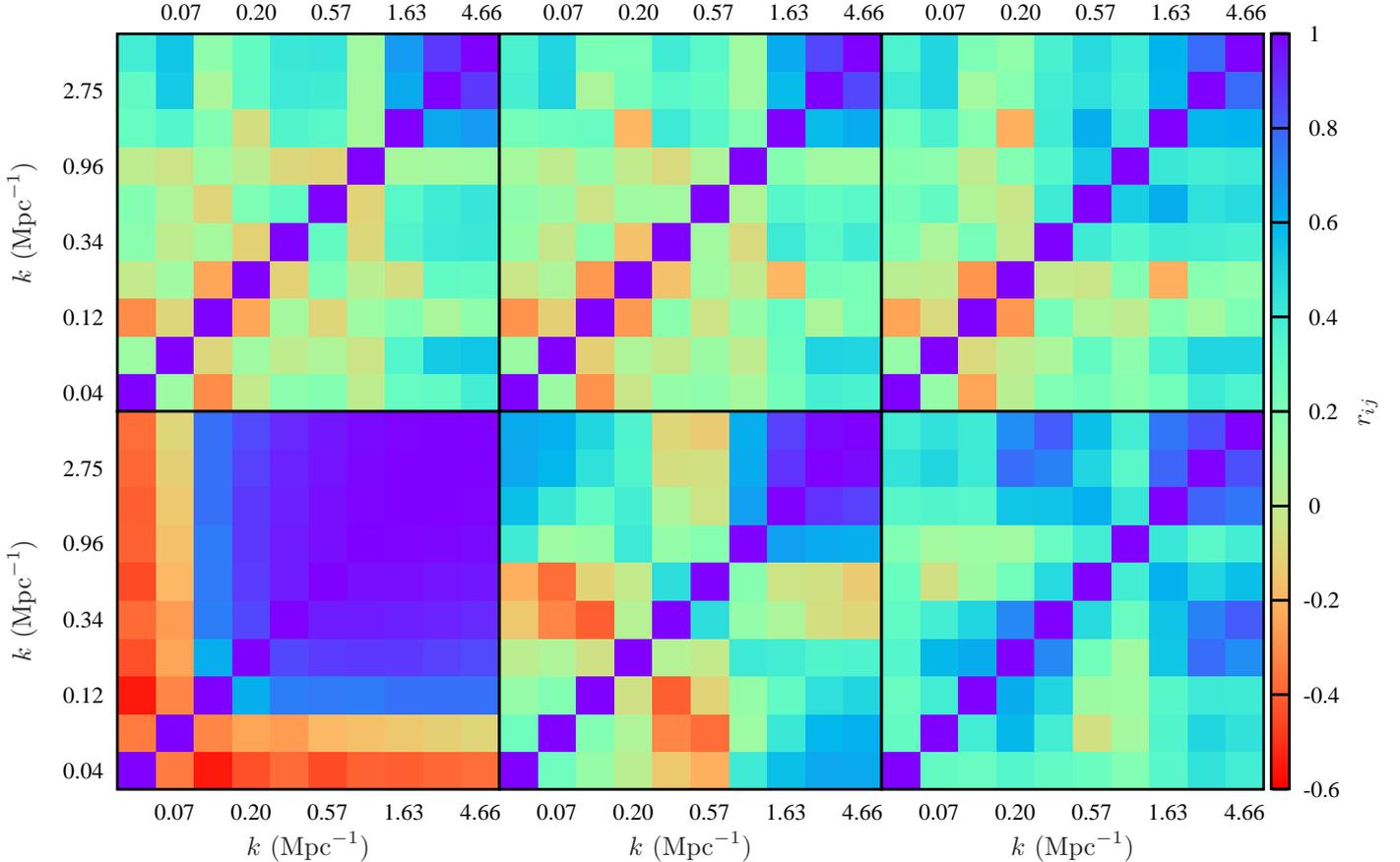}
\caption{This shows the correlation coefficient $r_{ij}$ estimated
  from SE for the values of $\xb$ listed in
  \autoref{tab:history}. Starting from top-left panel, the value of
  $\xb$ decreases clockwise.}
\label{fig:all_cov}
\end{figure*}

\begin{figure*}
\psfrag{k}[c][c][1][0]{{\large $k\, \, ({\rm Mpc}^{-1})$}}
\psfrag{cov}[c][c][1][0]{{\Large $r_{ij}$}}

\psfrag{z=13, xh1=0.98}[c][c][1][0]{{\large $\xb=0.98$}}
\psfrag{z=11, xh1=0.93}[c][c][1][0]{{\large $\xb=0.93$}}
\psfrag{z=10, xh1=0.86}[c][c][1][0]{{\large $\xb=0.86$}}
\psfrag{z=9, xh1=0.73}[c][c][1][0]{{\large $\xb=0.73$}}
\psfrag{z=8, xh1=0.50}[c][c][1][0]{{\large $\xb=0.50$}}
\psfrag{z=7, xh1=0.15}[c][c][1][0]{{\large $\xb=0.15$}}

\centering
\includegraphics[width=0.97\textwidth]{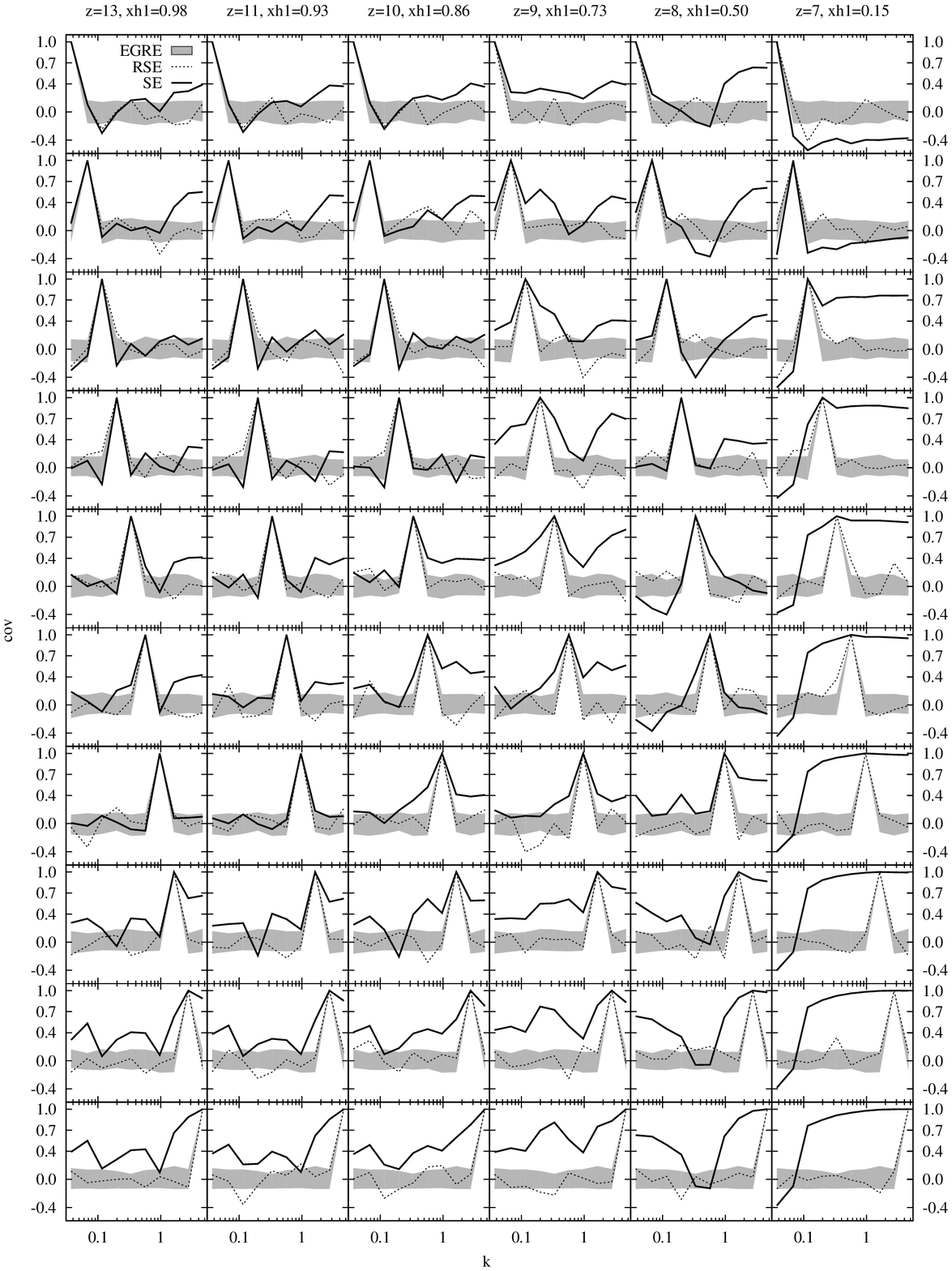}
\caption{This shows $r_{ij}$ estimated from SE (solid) and RSE
  (dotted). The shaded region represents the $[\delta r_{ij}]_{\rm G}$
  which quantifies the $1-\sigma$ fluctuations of the off-diagonal
  terms around $[r_{ij}]_{\rm G} = 0$ expected for a Gaussian random
  field.}
\label{fig:all}
\end{figure*}

%
We next consider the off-diagonal elements of the error-covariance matrix 
$\cov_{ij}$. These quantify the correlation between the errors in the power 
spectrum estimated at different $k$ bins. Following \p1, we use the 
dimensionless correlation coefficient 
\begin{equation}
r_{ij} = \frac{\cov_{ij}}{\sqrt{\cov_{ii}\,\cov_{jj}}}
\label{eq:corr}
\end{equation}
to quantify the off-diagonal elements of $\cov_{ij}$. It follows from the 
definition that $r_{ii}= 1$ for all the diagonal elements. The values of 
$r_{ij}$ lie in the range $-1 \le r_{ij} \le 1$, the errors in the $i~$th and 
$j~$th bins are completely correlated and anti-correlated if $r_{ij} = 1$ and 
$-1$ respectively. The errors in these two bins are uncorrelated if $r_{ij}=0$, 
and  intermediate positive or negative values ($-1 < r_{ij} < 1$) indicate 
partial correlations or partial anti-correlations respectively. We expect all 
the off-diagonal elements to have zero values $(r_{ij}=0)$ if the EoR 21-cm 
signal is a Gaussian random field. 
\autoref{fig:all_cov} shows $r_{ij}$ estimated from  SE for all the six $z$ 
and $\xb$ values listed in \autoref{tab:history}. We see that the errors in 
the three largest $k$ bins are quite strongly correlated even at the earliest 
stage $(\xb =0.98)$. The extent of this correlated region increases at the 
later stages of  reionization. We see that at $\xb=0.15$ the errors in all 
the $k$ bins barring the two smallest $k$ values are strongly correlated, 
the errors in these two smallest $k$ bins are anti-correlated with the errors 
in all the other bins. It is insightful (see \p1) to consider \autoref{fig:all} 
where each panel corresponds to a fixed value of $i$ for which it shows 
$r_{ij}$ as a function of $k_j$. Note that in all cases we have $r_{ij}=1$ for 
the diagonal terms which have $j=i$. Here  each horizontal set of panels 
corresponds to a fixed value of $i$ and each vertical set of panels 
corresponds to the fixed  value of $\xb$ shown at the top of the figure. 
%

%
\autoref{fig:all} shows the $r_{ij}$ values estimated from SE for all the six 
$z$ and $\xb$ values listed in \autoref{tab:history}. Following \p1, we have 
used the EGRE (Section~\ref{sec:EGRE}) to estimate $[\delta r_{ij}]_{\rm G}$ 
which provides an estimate of the fluctuation of the off-diagonal terms around 
$[r_{ij}]_{\rm G} = 0$ expected for a Gaussian random field. We have also shown 
$r_{ij}$ estimated from RSE (Section~\ref{sec:simRSE}). The RSE has been 
constructed with the intention of removing any correlation between the signal 
at different Fourier modes, and we see that the off-diagonal elements of 
$[r_{ij}]_{\rm  RSE}$ are nearly always within the shaded region corresponding 
to $[\delta r_{ij}]_{\rm G}$.
We first consider the results for $\xb = 0.98$ when the \HI essentially 
traces the underlying matter distribution. We have already seen 
(\autoref{fig:cov_SE_RSE}) that, even at this early stage, the error-variance 
for the three largest $k$ bins ($k > 1.1\,{\rm Mpc}^{-1}$) are 
significantly in excess of those predicted for a Gaussian random field. 
Here (Figures~\ref{fig:all_cov} and \ref{fig:all}) we see  that  errors 
in these three $k$ bins are not independent but they are highly correlated. 
We further see (\autoref{fig:all}) that the errors in most of the smaller 
$k$ bins also  are  correlated with the errors in the three largest $k$ 
bins. Two of the $k$ bins, $k \approx 0.1 \, {\rm Mpc}^{-1}$ and 
$k \approx 1.0\, {\rm Mpc}^{-1}$, however are exceptions, and the errors 
in these two $k$ bins are uncorrelated  to the errors in any of the other 
$k$ bins. The results do not change much for $\xb=0.93$, and we may 
interpret the results at the two highest $\xb$ values as reflecting the 
intrinsic properties of the underlying matter distribution. As mentioned 
earlier, the evolution of the underlying matter distribution is non-linear 
at small scales. This introduces non-Gaussianities that not only affect 
the error estimates at small scales but also causes the errors at large 
length scales to become correlated with the errors at small scales. The 
results are somewhat different when the neutral fraction falls further to 
$\xb=0.86$, and we may interpret this as the interplay of the intrinsic 
non-linearities of the underlying matter distribution and the ionization 
of the \HI which preferentially wipes out the 21-cm signal from the most 
non-linear regions. The extent of the small-scale correlation increases 
and we now find that the errors in the six largest $k$ bins are correlated. 
The errors in the two subsequent $k$ bins ($\approx 0.1$ and 
$0.2 \, {\rm Mpc}^{-1}$) are uncorrelated to the errors in any of the other 
bins, however the errors in the two smallest $k$ bins are correlated with 
the errors at the largest $k$ bins.
We next discuss the results for $\xb = 0.73$ where the reionization has 
extended further. Here (Figures~\ref{fig:all_cov} and \ref{fig:all}) we 
see that  errors in the three largest $k$ bins are strongly correlated 
among themselves, they are also correlated with the errors in all the 
other $k$ bins. Interestingly, the errors in even the smallest $k$ bins 
seem to be weakly correlated among themselves (\autoref{fig:all}). The 
results change somewhat for $\xb = 0.5$. The errors at the large $k$ 
modes again are highly correlated among themselves and also with the 
errors at the smallest $k$ bins. We have a  new feature here that the 
errors in the two bins with $k \approx 0.07$ and $0.12 \, {\rm Mpc}^{-1}$ 
are anti-correlated with the errors in the two bins with $k \approx 0.34$ 
and $0.57 \, {\rm Mpc}^{-1}$. We note that a similar weak anti-correlation 
has also been reported earlier in \p1 where the analysis was entirely 
restricted to $\xb = 0.5$. At the final stages of reionization 
($\xb \approx 0.15$) the errors in nearly all the $k$ bins, barring the 
two smallest $k$ values, are highly correlated (Figures~\ref{fig:all_cov} 
and \ref{fig:all}). The errors in the two smallest $k$ bins are weakly
anti-correlated with the errors in all the other $k$ bins. We see 
(\autoref{fig:pk}) that the 21-cm signal at this last stage of reionization 
originates from a few surviving low density \HI regions. This accounts for 
the highly correlated errors seen at this stage, however at present we do
not have an understanding of the source of the weak anti-correlation seen here.

\section{Summary and Discussion}
\label{sec:summary}
The detection of the EoR 21-cm signal is expected to become a 
reality with one or many of the current radio interferometers such 
as LOFAR, MWA and PAPER. It is also anticipated that the first 
detection would be carried out by measuring the 21-cm power spectrum. 
The error-covariance of the EoR 21-cm power spectrum is very important 
for predicting the prospects of a detection with these ongoing experiments. 
This is also equally important for quantifying the uncertainties in 
the EoR 21-cm power spectrum once it is measured by any of these 
experiments. Future instruments such as the SKA and HERA would be
able to measure the EoR 21-cm power spectrum with a better precision 
owing to their enhanced sensitivity. One of their major science goals 
is to use future high precision measurements to constrain different 
model parameters of reionization (e.g. \citealt{greig15,ewall16}). 
Earlier works which make predictions for different experiments have 
all assumed that the EoR 21-cm signal is a Gaussian random field. This 
implies that the error-variance of the 21-cm power spectrum in any $k$ 
bin depends only on the value of the power spectrum and the number of 
independent Fourier modes which contribute to the signal in that 
particular bin. This also implies that the errors in the different $k$ 
bins are uncorrelated. In our earlier work (\p1) we have used 
semi-numerical simulations to analyze the error-covariance matrix 
$\cov_{ij}$ expected at a particular redshift $z=8$ where the neutral 
fraction was assumed to have a value $\xb=0.5$. This reveals that the 
$\cov_{ij}$ estimated from $50$ independent realizations of the simulated 
EoR signal shows significant deviations from the predictions of a Gaussian 
random field. The effect of non-Gaussianity is expected to increase as 
reionization progresses. In this paper we have extended the analysis 
of \p1 to cover different stages of reionization.

We find (\autoref{fig:cov_SE_RSE}) that even at the very early stage
of reionization ($\xb = 0.98 $) the dimensionless error-variance 
 $[\dcov_{ii}]_{\rm SE}$ is $\sim 2$ -- $20$ times larger than the Gaussian
prediction  $[\dcov_{ii}]_{\rm RSE}$ at  intermediate and small length-scales
($k > 0.5\,{\rm Mpc}^{-1}$).  
The errors in the three largest $k$ bins ($k > 1.5\,{\rm Mpc}^{-1}$) 
are also highly correlated (Figures~\ref{fig:all_cov}). The error-variance 
is consistent with the Gaussian predictions  at large length-scales ($k \le
0.5\,{\rm   Mpc}^{-1}$).  We however find (\autoref{fig:all_cov} and
\ref{fig:all}) that the errors at large length-scales ($k < 0.1 \,{\rm
  Mpc}^{-1}$) and intermediate length-scales ($0.2 \le k \le 
0.6  \,{\rm Mpc}^{-1}$) 
are correlated with the errors in the three largest
$k$ bins ($k > 1.5\,{\rm Mpc}^{-1}$). As reionization proceeds, the results
show a similar behaviour at $\xb =0.93$. The results are slightly different
when the neutral fraction drops to $\xb =0.86$. The extent of the correlation
at small scales now extends to the $6$ largest $k$ bins. The strength of this
correlation, however, is slightly weakened compared to the earlier stages of
reionization. The error-variance at large length-scales ($k < 0.3 \,{\rm
  Mpc}^{-1}$) is consistent with the Gaussian predictions, but these errors 
still continue to be weakly correlated with the errors in the
largest $k$ bins. 

As reionization proceeds further  we find  that the peak value of
$[\dcov_{ii}]_{\rm   SE}$ drops from $\sim 80$ at $\xb=0.98$ to  $\sim 50$  at
$\xb=0.73$.  
However the length-scales across which the value of $[\dcov_{ii}]_{\rm SE}$
exceeds the Gaussian predictions increases to cover the range $k > 0.1 \,
{\rm Mpc}^{-1}$,   
possibly extending to the smallest $k$ bin $k  \ge 0.04 \, {\rm Mpc}^{-1}$.  
The errors in the $3$ largest $k$ bins are highly correlated among
themselves, and the errors in all the smaller $k$ bins are also correlated
with the errors in the $3$ largest  $k$ bins. We however do not find any
correlation between the errors at intermediate and large length-scales. 
The behaviour is a little complicated when the neutral fraction drops to
$\xb=0.5$. At small length scales  $k \ge 1.0 \, {\rm Mpc}^{-1}$,  
the value of $[\dcov_{ii}]_{\rm SE}$ increases relative to $\xb=0.73$, 
with the peak value of $[\dcov_{ii}]_{\rm  SE} \approx 500$.
The errors in the $4$ largest $k$ bins are also highly correlated. The
value of $[\dcov_{ii}]_{\rm SE}$, however, decreases relative to $\xb=0.73$
at intermediate length-scales $0.1 < k < 0.6 \, {\rm Mpc}^{-1}$. The errors
here are uncorrelated with those in the larger $k$ bins, however  they are 
anti-correlated with the error at $3$ smallest $k$ bins $k < 0.2 \, {\rm
Mpc}^{-1}$.  The errors in the smallest $k$ bins ($k< 0.2 \, {\rm Mpc}^{-1}$)
are correlated with those in the $3$ largest $k$ bins ($k > 1.0 \, {\rm
Mpc}^{-1}$) and anticorrelated with the errors at intermediate scales. 
At the final stage  ($\xb=0.15$)  we find that the  values of
$[\dcov_{ii}]_{\rm SE}$  ($\sim 3 \times 10^4$) increase considerably
relative to the earlier stages of reionization, and
exceeds the Gaussian predictions in  
all the bins above $k \ge0.1 \, {\rm Mpc}^{-1}$ (\autoref{fig:cov_SE_RSE}).
The errors in the large $k$  bins are also highly correlated
(\autoref{fig:all_cov}).  
The error-variance in the two smallest $k$ bins are consistent with the
Gaussian predictions, however the errors here  are anti-correlated with the
errors in all the larger $k$ bins.

The value of $[\dcov_{ii}]_{\rm SE}$ peaks at the largest $k$
bin $(4.66 \, {\rm Mpc}^{-1})$ through all stages of reionization. It is
interesting to note that this 
peak value decreases from $\sim 80$ at $\xb=0.98$ to $\sim 50$ at $\xb=0.73$ 
in the early stages of reionization.  The peak value subsequently increases
sharply to $3 \times 10^4$ at $\xb=0.15$. A similar behaviour is seen at $k=
2.75 \, {\rm Mpc}^{-1}$ (right panel of \autoref{fig:cov_SE}) which is the
second largest $k$ bin. In contrast, the value of
$[\dcov_{ii}]_{\rm SE}$ at the 
intermediate length-scales $k \sim 0.5 \, {\rm Mpc}^{-1}$ initially decreases
slightly  as $\xb$ falls from $0.98$ to $0.93$, and then increases to
$[\dcov_{ii}]_{\rm SE}\sim 10$ at $\xb=0.73$ (middle panel of
\autoref{fig:cov_SE}). The value of
$[\dcov_{ii}]_{\rm SE}$  subsequently dips to $\sim 3$ for $\xb=0.5$ and
finally increases sharply to $\sim 100$ at $\xb=0.15$.  We find a similar
behaviour at large length-scales (left panel of \autoref{fig:cov_SE}). Here
the peak values of $[\dcov_{ii}]_{\rm SE}$ are in the range $4$ to $5$, and it 
is not very clear if these are  significantly in excess of the Gaussian 
predictions. It is important to note that even though the error-variance is
consistent with the Gaussian predictions at large length-scales, the errors
here are correlated with those at smaller length-scales. This clearly
indicates that the non-Gaussian effects are important even at the largest
length-scales. 
The analysis of this paper also enables us to estimate the trispectrum of the
EoR 21-cm signal $t_{ii}=[\dcov_{ii}]_{\rm SE} - [\dcov_{ii}]_{\rm RSE}$
(equation~\ref{eq:tii}), however we have not considered this here.  

The \HI traces the underlying matter distribution during  
the earliest stage of reionization ($\xb=0.98$). The non-Gaussianities here
arise  due to the small scale non-linear gravitational clustering of the
underlying matter distribution. We can interpret the excess error-variance at
small length-scales as arising from this non-linear gravitational clustering.
Further, this also influences large length-scales through mode coupling
\citep{bharadwaj96b,bharadwaj96}, and provides an interpretation for the
correlation between the errors at large and intermediate length-scales with
those at small scales. It is interesting to note that a similar effect is
important in the context of galaxy redshift surveys aiming to measure the 
Baryon Acoustics Oscillations (BAO) where it is found that that the errors at
the BAO length-scales ($\sim 150 \, {\rm Mpc}$) are influenced by the small
scale non-linear gravitational clustering
\citep{hamilton06,neyrick08,neyrinck11,Harnois13,mohammed14,caron14}.

As reionization proceeds, the \HI in the non-linear  high density peaks is
preferentially ionized in the inside-out reionization scenario implemented in
our simulations, resulting  in a drop in the non-Gaussianity of the  
21-cm signal. At small scales this  causes the error-variance to decrease as
$\xb$ falls from $0.98$ to $0.86$.  This also causes a weakening of the
strength of the correlation between the errors in the different large $k$
bins. In the subsequent two stages of reionization 
the 21-cm signal is dominated by the  discrete ionized bubbles
(\autoref{fig:HI_map}) which have  diameters spread around $10-20 \, {\rm
  Mpc}$ and  $40-50 \, {\rm Mpc}$ for $\xb=0.73$ and $0.5$ respectively.  
We expect the Poisson fluctuations from these discrete bubbles to 
contribute significantly to the non-Gaussianity  \citep{bharadwaj05a}, and we
may interpret the excess error-variance here as arising from an interplay
between the contribution from the discrete bubbles and the intrinsic
non-Gaussianity of the underlying matter distribution.  At the final stage
($\xb=0.15$) the 21-cm signal  is dominated by a few
discrete surviving \HI regions. The excess error-variance and the correlation
between the errors here may be interpreted as arising from the Poisson
fluctuations due to these discrete \HI regions which have sizes  in the
range $30-50 \, {\rm Mpc}$ (\autoref{fig:HI_map}). At present we do not have
an interpretation for  the anti-correlation seen at the last two stages of
reionization. 

Our entire analysis is based on rather simple semi-numerical
simulations where the reionization history is determined
by three parameters namely $M_{\rm min}$, $N_{\rm ion}$ and $R_{\rm mfp}$
(defined in subsection  \ref{sec:simSE}). The values of these parameters
can be tuned to obtain different reionization histories for which the
predicted 21-cm signal also differs. These simulations also 
do not incorporate X-ray heating
(e.g. \citealt{ghara14,greig15, mesinger16}) which will possibly
introduce additional non-Gaussianity at the early stages. The 
predictions will possibly also differ if one includes fully
coupled 3D radiative transfer simulations (e.g. \citealt{gnedin16})
or if one considers variations in the reionization sources
(e.g. \citealt{majumdar16}). One may question as to how dependent 
our conclusions are on the parameter values and the method of 
simulation.
To address this issue  we have repeated the analysis  using the publicly
available semi-numerical code   
21cmFAST \citep{mesinger11}. Like our semi-numerical scheme, 21cmFAST 
also has three parameters $(T_{\rm  vir,min},\, \zeta_{\rm ion},\, R_{\rm
  mfp})$ which are roughly equivalent to the parameters 
($M_{\rm min},N_{\rm  ion},R_{\rm mfp}$) used in our simulations. The values
of the 21cmFAST parameters were chosen so as to achieve 
$\xb=0.5$  at $z=8.0$, and the results were compared with our predictions at
$z=8$. The results from  21cmFAST and a detailed 
comparison with our predictions is presented  in Appendix~\ref{sec:com}.  To
summarize the findings, the 21-cm maps (\autoref{fig:com_map}), the power
spectrum and the error-variance (\autoref{fig:com_pk_cov_SE})
are in good agreement at large length-scales ($ k \le 0.2 \, {\rm
  Mpc}^{-1}$). The power spectrum and the error-variance from 21cmFAST  
exceeds our predictions at larger $k$, the two are however qualitatively
similar. In contrast, the correlation between the errors in the  different $k$
bins (\autoref{fig:com_cov}) are quite different for 21cmFAST as compared to
our predictions. We see that for 21cmFAST the errors at large length-scales  
($ k \le 0.2 \, {\rm  Mpc}^{-1}$) are strongly anti-correlated with those at
small scales ($ k \ge 0.3 \, {\rm  Mpc}^{-1}$), a feature which is not seen in
our predictions. In conclusion of this comparison we note that 
the error predictions presented here do depend on the
method and the parameters of the simulation. However, 
we can definitely treat our predictions as being indicative of  the 
magnitude and the qualitative nature of the deviations from the Gaussian
predictions at different stages of reionization. We also note that 
\citet{majumdar14} presents a detailed comparison of the simulation
methods implemented in 21cmFAST, our simulations and also a radiative transfer
code C$^2$-RAY.  

The non-Gaussian nature of the EoR 21-cm signal plays a very significant role in
determining the error statistics of the power spectrum, and it is very
important to include this when making predictions for ongoing and future
experiments. This will also be important for interpreting the results once a
detection is made. The various predictions till now have all assumed the
EoR 21-cm signal to be a Gaussian random field whereby the power spectrum
error-covariance matrix is diagonal and can be easily calculated if the power
spectrum is known. In reality, as shown in this work,  we expect the
error-covariance to be non-diagonal (e.g. \citealt{liu14, liu14a}) and
depend on both the power spectrum and
the trispectrum. The most straight forward approach would be to use a
statistical ensemble of the simulated 21-cm signal to estimate the
error-covariance matrix, as done here, and  use this to make more realistic
predictions for future and ongoing experiments. It may be noted that we plan
to do this in future work. The main difficulty here is that the reionization
history and the underlying 21-cm signal are largely unknown. Further, the
error predictions also depend on the parameters and the method of simulation.
Consequently, the error predictions are inherently uncertain due to the lack
of our knowledge of the reionization process and our limited ability to
model reionization. However, in 
all cases we expect the actual error predictions to exceed the Gaussian
predictions and we can treat the Gaussian predictions as being upper limits to
the signal to noise ratio (SNR) that can be achieved in any experiment. The
inclusion of non-Gaussianity will degrade the SNR relative to the Gaussian
predictions, the non-Gaussian prediction will however vary with reionization
model and method of simulation. The best one can do with the limited
understanding of reionization available at present is to have error
predictions for different models and simulation methods. With the advent of
detections one expects to narrow down the uncertainty in the reionization
history and the 21-cm signal.  This in turn should lead to better error
predictions which  should result in a  better interpretation 
of the detection signal. 

In future work  we plan to use the reionization model and simulation method
of this paper to make predictions for ongoing and future
EoR experiments including the system noise and  other telescope specific
effects.


\section*{Acknowledgements}
Authors would like to acknowledge the use of the publicly available
21cmFAST code \citep{mesinger11} for a part of the analysis demonstrated
in this paper. RM would like to thank Kanan K. Datta for useful
discussions. SM would like to acknowledge the financial assistance
from the European Research Council under ERC grant number
638743-FIRSTDAWN and from the European Unions Seventh Framework
Programme FP7-PEOPLE-2012-CIG grant number 321933-21ALPHA.


\bibliographystyle{mnras} 
\bibliography{refs}


\appendix
\section{Comparison of different models}
\label{sec:com}
To test whether our conclusions  depend on the parameter values and
the method of simulation, we have repeated the  analysis using a
publicly available semi-numerical code 21cmFAST \citep{mesinger11}.
The values of the 21cmFAST parameters $(T_{\rm  vir,min},\, \zeta_{\rm ion},\,
R_{\rm mfp})$ were chosen to be $(3\times10^4\,{\rm K},\, 15,\, 20\,{\rm Mpc})$
so as to achieve $\xb=0.5$  at $z=8.0$, and the results were compared
with our predictions at a single redshift $z=8$. The spatial resolution
$0.56\,{\rm Mpc}$ and the simulation volume $V=[215\, {\rm Mpc}]^3$ are
maintained the same for both methods.

The two different simulation methods  being compared here  mainly differ on
three counts: (1.) Our method uses a cosmological $N$-body simulation to
calculate the matter density field at the desired redshift whereas 21cmFAST
uses Zel'dovich approximation (2.) Our method uses FoF to identify the halos
which host the ionizing sources whereas 21cmFAST uses a conditional
Press-Schechter formalism to calculate the total mass in collapsed halos at
every grid point (3.) The particles which represent the \HI distribution are
displaced along the line of sight to incorporate the effect of peculiar
velocities whereas 21cmFAST implements this by multiplying the brightness
temperature map with a factor involving the derivative of the radial component
of peculiar velocity on a fixed grid without actually moving the \ion{H}{I}. 
\citet{majumdar14} presents a detailed comparison of the simulation
methods implemented in 21cmFAST and in  our simulations.

\begin{figure}
\centering
\includegraphics[width=0.47\textwidth, angle=0]{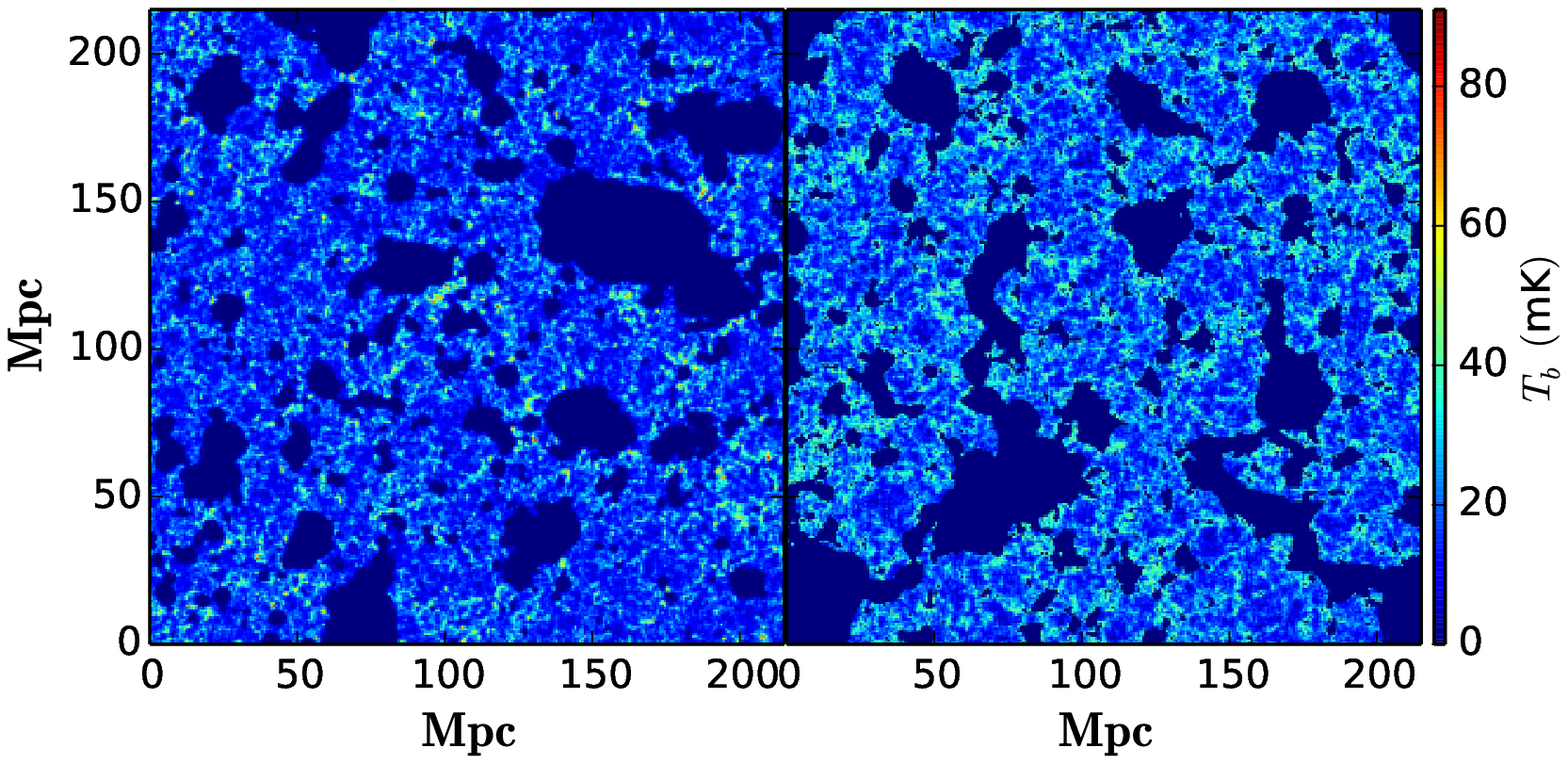}
\caption{This shows two dimensional sections through 
\HI 21-cm brightness temperature maps simulated  using our 
method (left panel) and 21cmFAST (right panel).}
\label{fig:com_map}
\end{figure}

\autoref{fig:com_map} provides a visual comparison of the \HI 21-cm
brightness temperature maps for the two different simulation methods.
A visual inspection confirms that the maps look roughly similar in
nature. However, there are some differences in the ionized bubble sizes
and the small scale \HI structures. These differences are caused by the
differences in the simulation techniques discussed above.
%
%
%

\begin{figure}
\psfrag{cov}[c][c][1][0]{\small $\dcov_{ii}$}
\psfrag{pk}[c][c][1][0]{\small ${\Delta^2_{\rm b}}\, \, ({\rm mK})^2$}
\psfrag{k}[c][c][1][0]{\small $k\,({\rm Mpc}^{-1})$}

\centering
\includegraphics[width=0.47\textwidth]{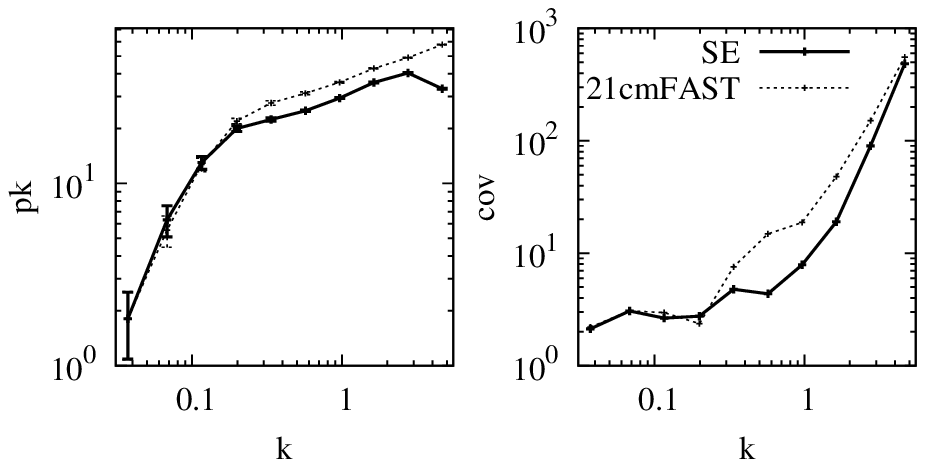}
\caption{This shows the dimensionless  21-cm brightness temperature
fluctuations $\Delta_{\rm b}^2(k)$ (left panel) and the dimensionless
variance  $\dcov_{ii}$ as a function of $k$ (right panel). The predictions
of our simulation (solid) are compared with those from 21cmFAST (dashed).}
\label{fig:com_pk_cov_SE}
\end{figure}

%
We have used $50$ realizations of the 21cmFAST simulations to estimate the
mean power spectrum and the error-covariance matrix. 
The left panel of \autoref{fig:com_pk_cov_SE} shows a  comparison
of $\Delta_{\rm b}^2(k)$ calculated using the two different methods. 
We observe that the power spectrum is in good
agreement at large length scales ($k \le 0.2\,{\rm Mpc}^{-1}$), 
the results from 21cmFAST  are however in excess of our predictions 
at smaller length scales. 
The right panel of \autoref{fig:com_pk_cov_SE} shows a comparison of
the dimensionless error-variance $\dcov_{ii}$.  Here also we find that the two
methods are consistent at  large length scales ($k \le 0.2\,{\rm Mpc}^{-1}$)
whereas the results from 21cmFAST  are however in excess of our predictions 
at smaller length scales.
%

\begin{figure}
\psfrag{k}[c][c][1][0]{\small $k\, \, ({\rm Mpc}^{-1})$}
\psfrag{cov}[c][c][1][0]{\small $r_{ij}$}

\centering
\includegraphics[width=.47\textwidth]{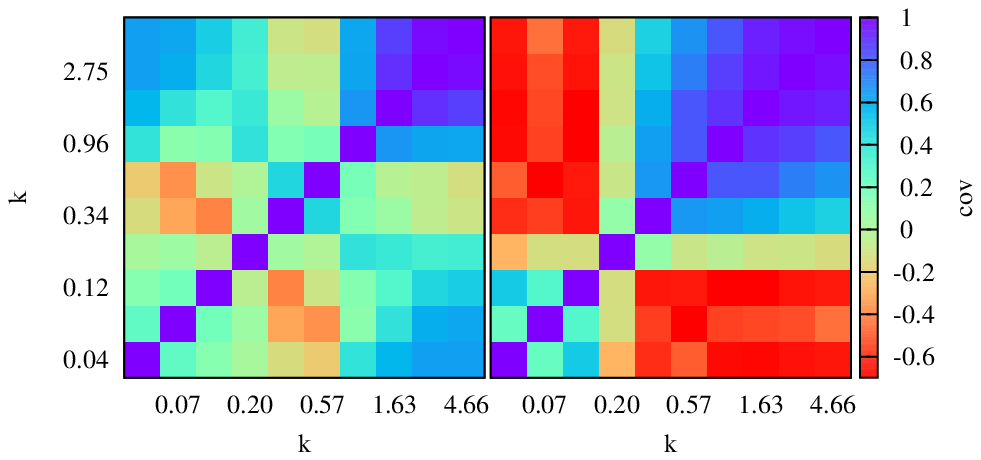}
\caption{This shows the correlation coefficient $r_{ij}$ estimated using our
  simulations (left panel) and 21cmFAST (right panel).}
\label{fig:com_cov}
\end{figure}

%

\autoref{fig:com_cov} shows a comparison of  the correlation coefficient
$r_{ij}$ estimated from the two different methods. We observe that the errors
in the largest $k$ bins are strongly correlated for both models. However, the
extent of this correlated region is  larger for 21cmFAST
($k \ge 0.3\,{\rm Mpc}^{-1}$)   as compared to our predictions 
($k \ge 1.0\,{\rm Mpc}^{-1}$). 
 We further observe that the errors in the four smallest $k$
bins ($k \le 0.2\,{\rm Mpc}^{-1}$) are strongly anti-correlated with those at
small scales ($ k \ge 0.3 \, {\rm  Mpc}^{-1}$) for 21cmFAST. This feature is
not seen in our predictions where the corresponding errors are found to be
weakly correlated. 

In conclusion of this comparison we note that the error predictions
presented here do depend on the method and the parameters of the simulation.

\bsp
\label{lastpage}
\end{document}